\documentclass[9pt]{article}
\usepackage{spconf,amsmath,amsfonts,graphicx,algorithm,algpseudocode,epstopdf,multirow,threeparttable,caption,subcaption,multicol,subcaption,balance, mathtools,cite}

\title{A Two-Step Approach for Narrowband Source Localization in Reverberant Rooms}
\name{\fontsize{11}{13}\selectfont Wei-Ting Lai, Lachlan Birnie, Thushara Abhayapala, Amy Bastine, Shaoheng Xu, Prasanga Samarasinghe}
\address{\fontsize{11}{13}\selectfont Audio and Acoustic Signal Processing Group, The Australian National University, Canberra, Australia}

\begin{document}
%
\maketitle
\begin{abstract}
This paper presents a two-step approach for narrowband source localization within reverberant rooms. The first step involves dereverberation by modeling the homogeneous component of the sound field by an equivalent decomposition of planewaves using Iteratively Reweighted Least Squares (IRLS), while the second step focuses on source localization by modeling the dereverberated component as a sparse representation of point-source distribution using Orthogonal Matching Pursuit (OMP). The proposed method enhances localization accuracy with fewer measurements, particularly in environments with strong reverberation. A numerical simulation in a conference room scenario, using a uniform microphone array affixed to the wall, demonstrates real-world feasibility. Notably, the proposed method and microphone placement effectively localize sound sources within the 2D-horizontal plane without requiring prior knowledge of boundary conditions and room geometry, making it versatile for application in different room types.
\end{abstract}


\section{Introduction}
\label{sec:intro}

Sound source localization plays a crucial role in various acoustic applications, such as speech enhancement \cite{enhancement1,enhancement2}, source separation \cite{fahim,s-sh}, and sound field translation \cite{translation}. In environments with strong reverberation, the challenge of source localization experiences a considerable escalation. Several source localization methods have been applied to reverberant environments, such as beamforming \cite{beamforming}, MUSIC \cite{music_l10n_2,music_l10n}, SRP-PHAT \cite{srp-phat}, and CLEAN \cite{clean}. These methods use statistical properties of signals to estimate source positions. However, their performance declines significantly when processing narrowband sources and correlated reflections.

Recently, some sparsity-based methods, such as LASSO \cite{lasso}, Orthogonal Matching Pursuit (OMP) \cite{omp_l10n}, and Sparse Bayesian Learning \cite{sbl}, have been introduced to overcome the limitation of narrowband localization by assuming sparse distribution of sources in spatial domain. Nevertheless, these methods continue to struggle in strong reverberation due to the interference of room reflections. Overcoming this challenge often requires either prior knowledge of room geometry and boundary conditions \cite{prior} or the use of sound field decomposition \cite{decompose}. Sound field decomposition entails separating the sound field into a direct component and reverberant component and modeling room reflections as a sum of planewaves or spherical-harmonics \cite{vekua2}. For source localization tasks, the latter approach models room reflections through a sum of planewaves and estimate the source positions based on the dereverberated sound field, such as wavefield separation projector processing (WSPP) \cite{wspp} and sparsity-based spherical harmonics model (S-SH) \cite{s-sh}. These approaches can handle a wide range of scenarios without requiring prior information. However, these methods demand a large number of microphones for modeling the reverberant component and are restricted by the geometry of the microphone array.

In this paper, we propose a similar two-step approach of dereverberation and sparsity based source localization that uses fewer microphones. We first dereverberate the sound field captured in a room by modeling the reverberant component as an equivalent planewave decomposition model \cite{vekua2}. The second step models the dereverberated sound field as superposition of the sparse point-sources and determine the source positions based on the sparse equivalent source method \cite{fernandez2017sparse}. We verify the proposed method through simulations in a conference room scenario, using a linear microphone array around the middle horizontal plane of the wall. The proposed placement effectively captures sound field information, while also not being constrained by room geometry. The results demonstrate that the two-step approach using separate algorithms improves source localization with fewer microphones, especially in rooms with strong reverberation.

\section{Problem Formulation}
\label{sec:formulation}

\begin{figure}[h]
  \centering
  \centerline{\includegraphics[width= 6.5 cm]{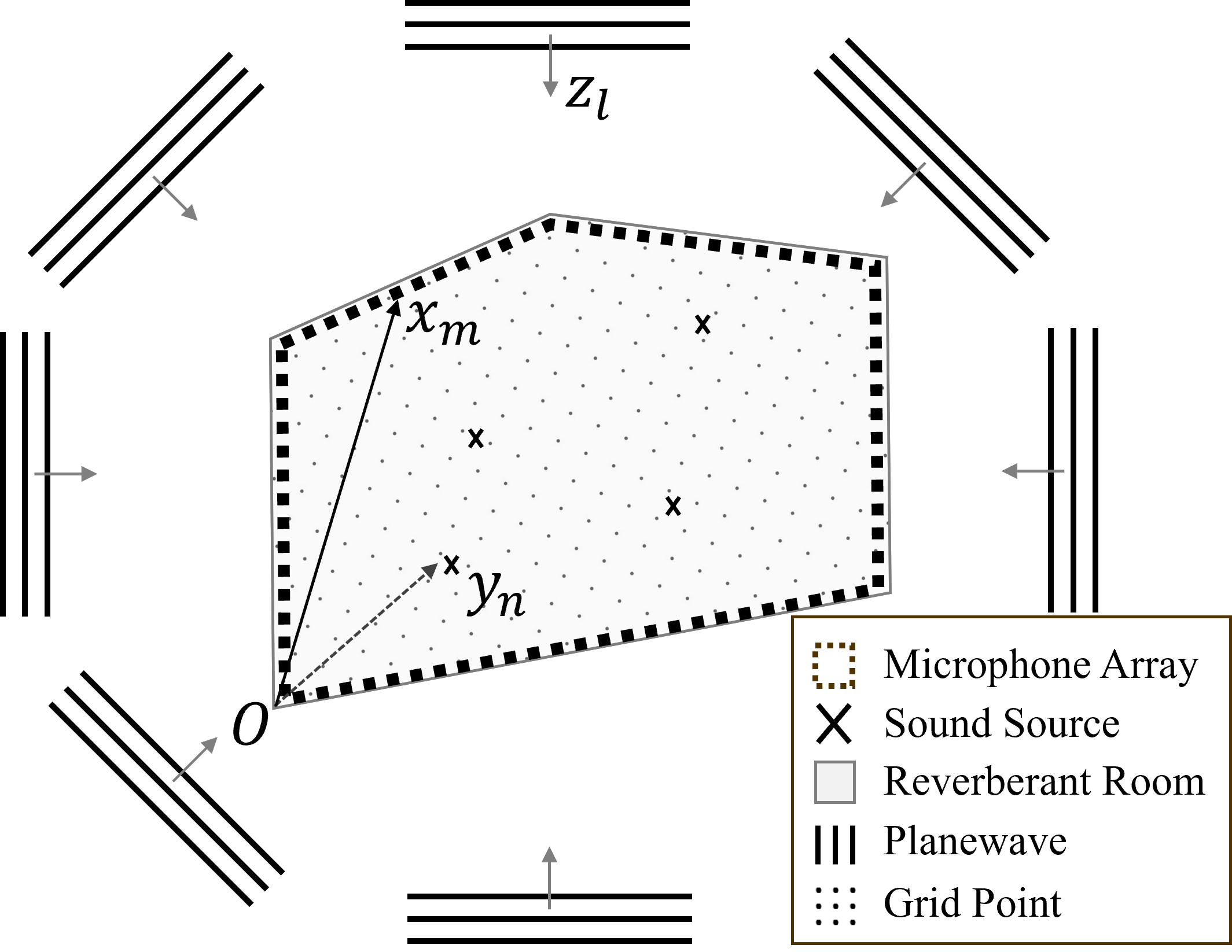}}
  \caption{Framework for the proposed method.}\medskip
  \vspace{-0.7cm}
\end{figure}

Consider $N$ sound sources in a reverberant room, along with $M$ microphones uniformly placed affixed to the walls, as illustrated in Fig. 1. The positions of the sound sources and microphones are defined as $\boldsymbol{y}_n \equiv (x_n, y_n, z_n)$ for $n=1,2,\dots,N$ and $\boldsymbol{x}_m \equiv (x_m, y_m, z_m)$ for $m=1,2,\dots,M$, respectively, with respect to the coordinate origin $O$ at the front-left-bottom corner of the room. Note, this configuration indicates that the $N$ sources are positioned inside the microphone array. 

The sound pressure received by the $m^{\text{th}}$ microphone is:
\begin{equation}
s(k, \boldsymbol{x}_m)= {\textstyle \sum_{n=1}^N} G(k, \boldsymbol{x}_m, \boldsymbol{y}_n) \alpha_n(k)
\end{equation}
where $k=2\pi f/c$ is the wave number, $f$ is frequency, $c$ is the speed of sound, $s(k, \boldsymbol{x}_m)$ represents the pressure, $G(k, \boldsymbol{x}_m, \boldsymbol{y}_n)$ denotes the room transfer function between the $n^{\text{th}}$ source and the $m^{\text{th}}$ microphone, and $\alpha_n(k)$ denotes the signal produced by the $n^{\text{th}}$ source. Note that this formulation does not include additive noise for brevity, which will be considered in the simulation.

Based on sound field decomposition \cite{decompose} and Vekua's theory \cite{vekua2,vekua}, any reverberant sound field can be partitioned into a sum of its particular and homogeneous solutions—equivalent to the direct and reverberant components, respectively. Within a bounded convex region, the reverberant sound field component can be well approximated using a finite number of planewave functions distributed over a spherical region. Consequently, the sound field can be expressed as a linear combination of $N$ direct-path point-source Green's function and $L$ planewave Green's function. Equation (1) can thus be decomposed as follows:
\begin{equation}
\begin{split}
& s(k{,}\boldsymbol{x}_m)
\approx 
\\ &
\underbrace{\sum_{n=1}^N G_0(k{,} \boldsymbol{x}_m{,} \boldsymbol{y}_n)
\alpha_n(k)}_\text{Direct}
{+}
\underbrace{\sum_{\ell=1}^L W(k{,} \boldsymbol{x}_m{,} \hat{\boldsymbol{z}}_\ell)
\beta_\ell(k)}_\text{Reverberant}
\end{split}
\end{equation}
where $G_0(k, \boldsymbol{x}_m, \boldsymbol{y}_n)=e^{i k\|\boldsymbol{x}_m-\boldsymbol{y}_n\|}/({4 \pi\|\boldsymbol{x}_m-\boldsymbol{y}_{n}\|})$ represents the direct-path Green's function between the $n^{\text{th}}$ source and the $m^{\text{th}}$ microphone in a free-field environment. $W (k, \boldsymbol{x}_m, \hat{\boldsymbol{z}}_\ell )=e^{-i k \hat{\boldsymbol{z}}_\ell \cdot \boldsymbol{x}_m}$ represents the $\ell^{\text{th}}$ planewave Green's function at $m^{\text{th}}$ microphone, with $\hat{\boldsymbol{z}}_\ell$ denoting the $\ell^{\text{th}}$ planewave's incident direction for $\ell=1,2,\dots,L$. The coefficient $\beta_\ell(k)$ represents the weight of the $\ell^{\,\text{th}}$ planewave.

In order to find the source positions, the direct component is modeled using a dictionary of $J$ point-sources within the room based on the sparse equivalent source method \cite{fernandez2017sparse}. This method assumes that sound sources exhibit quantity sparsity in spatial domain, indicating $N\ll J$. Finally, equation (2) can be reformulated as follows:
\begin{equation}
s(k{,} \boldsymbol{x}_m )
{\approx}
\sum_{j=1}^J G_0(k{,} \boldsymbol{x}_m{,} \boldsymbol{y}_j ) \alpha_j(k)
{+}
\sum_{l=1}^L W (k{,} \boldsymbol{x}_m{,} \hat{\boldsymbol{z}}_\ell ) \beta_\ell(k)
\end{equation}
where $G_0(k, \boldsymbol{x}_m, \boldsymbol{y}_j)$ denotes the free-field point-source Green's function between the $j^{\text{th}}$ grid point and the $m^{\text{th}}$ microphone. Hence, (3) is represented in matrix form as:
\begin{equation}
\boldsymbol{\mathbf{s}}=\mathbf{G}_0 \boldsymbol{\alpha}+\mathbf{W} \boldsymbol{\beta},
\end{equation}
where $\boldsymbol{\mathbf{s}} \in \mathbb{C}^{M}$ denotes the measured pressure from $M$ microphones, and $\mathbf{G_0} \in \mathbb{C}^{M \times J}$ and $\mathbf{W} \in \mathbb{C}^{M \times L}$ represent the dictionary matrices for point-sources and planewaves, respectively. The weight coefficient vectors for point-sources and planewaves are denoted as $\boldsymbol{\alpha} \in \mathbb{C}^{J}$ and $\boldsymbol{\beta} \in \mathbb{C}^{L}$.

Equation (4) is the sound field decomposition model in reverberant environments. As $L$ becomes sufficiently large, most elements of $\boldsymbol{\beta}$ can be approximated as nearly zero. Additionally, relying on the spatial sparsity assumption for sound source distribution, the $\boldsymbol{\alpha}$ vector tends to also have few non-zero elements, given that $N$ is significantly smaller than $J$. Therefore, the weight coefficients for point-sources $\boldsymbol{\alpha}$ and planewaves $\boldsymbol{\beta}$ can be determined through sparse optimization, as shown below \cite{s-sh}:
\begin{equation}
\underset{\boldsymbol{\alpha}, \boldsymbol{\beta}}{\operatorname{argmin}}\|\boldsymbol{\alpha}\|_1+\lambda\|\boldsymbol{\beta}\|_1     
\quad\text { s.t. } \mathbf{s}=\mathbf{G}_0 \boldsymbol{\alpha}+\mathbf{W} \boldsymbol{\beta},
\end{equation}
where $\lambda$ is a regularization term. 

The objective in this study is to estimate $\boldsymbol{\alpha}$ and determine the corresponding $N$ source positions $\boldsymbol{y}_n$ by solving (4), while assuming that the number of sources $N$ is known.
We propose a two-step process to solve (4) next.

\section{Two-Step Sparse Localization Method}
\label{sec:algorithm}
\subsection{Dereverberation}
\label{ssec:dereverberation}

For the first step, we estimate the reverberant component and perform dereverberation. We start by rearranging (4) as:
\begin{equation}
\boldsymbol{\mathbf{s}}= \boldsymbol{A}\boldsymbol{\gamma}
\end{equation}
where $\boldsymbol{A=[\mathrm{G}_0,\mathrm{W}]} {\in} \mathbb{C}^{M \times (J+L)}$, and $\boldsymbol{\gamma}=[\boldsymbol{\alpha},\boldsymbol{\beta}] {\in} \mathbb{C}^{(J+L)}$. In this context, we assume $M{<}(J{+}L)$, such that the number of measurements is fewer than the combined total modeled point-sources (grid points) and planewaves in the dictionary. Hence, the estimation of $\boldsymbol{\gamma}$ is an underdetermined problem.

We consider solving the linear regression problem (6) through Iteratively Reweighted Least Squares (IRLS) \cite{irls}, exploiting an $\ell^p$-norm approach by adding weights to $\ell^2$-norm optimization that iteratively refine the solution's sparsity (where $0<p\leq2$):
\begin{equation}
\min _\gamma {\textstyle \sum_{i=1}^{\mathcal{L}}} w_i \boldsymbol{\gamma}_i^2, \quad \text { subject to } \mathbf{s}=A \boldsymbol{\gamma}
\end{equation}
where $w_i= |\boldsymbol{\gamma}_i^{(v-1)} |^{p-2}$ are the weights computed from the previous iteration $\boldsymbol{\gamma}^{(v-1)}$. Hence, this iterative optimization is a $\ell^p$ objective function. The next iteration $\boldsymbol{\gamma}^{(v)}$ is as follows:
\begin{equation}
\boldsymbol{\gamma}^{(v)}=Q_v A^T (A Q_v A^T )^{-1} \mathbf{s}
\end{equation}
where $Q_v$ is the diagonal matrix with $1/w_i= |\boldsymbol{\gamma}_i^{(v-1)} |^{2-p}$.  We obtain the reverberant component of the sound field from the estimated weight coefficients of planewaves $\boldsymbol{\hat{\beta}}$, which is extracted from $\boldsymbol{\hat{\gamma}}$. Therefore, the dereverberation process can be represented as:
\begin{equation}
\boldsymbol{\hat{\mathbf{s}}}_0 = \boldsymbol{\mathbf{s}} - \mathbf{W} \boldsymbol{\hat{\beta}}
\end{equation}
where $\boldsymbol{\hat{\mathbf{s}}}_0$ denotes the estimated direct component, which we use as the dereverberated sound field component in the subsequent source localization step.

\subsection{Source Localization}
\label{ssec:localization}

For the second step, we re-estimate $\boldsymbol{\alpha}$ from the dereverberated component using OMP. Given the spatial sparsity assumption for sound source distribution, it follows that $\boldsymbol{\alpha}$ is assumed to have $N$ non-zero elements. Hence, we propose using OMP. OMP is a greedy algorithm, adapting an iterative process to select the most relevant sound source at each step \cite{omp}. By forcing inactive weights to zero, OMP improves the accuracy of source localization estimation. Utilizing the result from (9), we express the direct sound field component as:
\begin{equation}
\boldsymbol{\hat{\mathbf{s}}}_0 = \mathbf{G}_0 \boldsymbol{\widetilde{\alpha}}
\end{equation}
where $\boldsymbol{\widetilde{\alpha}}$ denotes the estimated point-source weight vector determined by OMP. In practice, source positions are found by selecting the weight coefficient with the highest correlation in each iteration. The OMP algorithm as follows:
\begin{algorithm} [H]
        \caption{Dereverberated OMP localization}
        \label{alg:ALG1}
        \textbf{Input:} measurements $\hat{\mathbf{s}}_0$, point-source dictionary $\mathbf{G}_0$, number of sources $N$\\
        \textbf{Output:} estimated weight coefficients $\boldsymbol{\widetilde{\alpha}}$, estimated source positions $\boldsymbol{\widetilde{y}_n}$
        \begin{algorithmic}
            \State $\Lambda_0=\o, \Psi_0=\o, \boldsymbol{g_i} = \mathbf{G}_0[:,i]$ for $i \in \{1, 2, \ldots, J\}$
            \For {  $n = 1$ to $N$ }
            \State $\boldsymbol{\widetilde{y}_n} \leftarrow\arg \max _{j \in\{1, 2, \ldots, J\}} | \boldsymbol{\langle\hat{\mathbf{s}}_0}, \boldsymbol{g_j} \rangle |$
            \State $\Psi_n \leftarrow\Psi_{n-1} \cup \{\boldsymbol{\widetilde{y}_n} \}$
            \State $\Lambda_n \leftarrow\Lambda_{n-1} \cup \{\boldsymbol{g_j} \}$
            
            \State $\boldsymbol{\hat{\mathbf{s}}}_0  \leftarrow \boldsymbol{\hat{\mathbf{s}}}_0-\Lambda_n \Lambda_n^{\dagger} \boldsymbol{\hat{\mathbf{s}}}_0$
            \EndFor
            \State $\widetilde{\alpha}_{\Psi_N}  \leftarrow \Lambda_N \Lambda_N^{\dagger} \boldsymbol{\hat{\mathbf{s}}}_0$
        \end{algorithmic}
    \end{algorithm}

\section{SIMULATION RESULTS}
\label{sec:result}

\begin{figure*}[ht]
\hspace{-0.8cm}
  \begin{subfigure}[b]{0.17\linewidth}
    \includegraphics[width=1.1\linewidth]{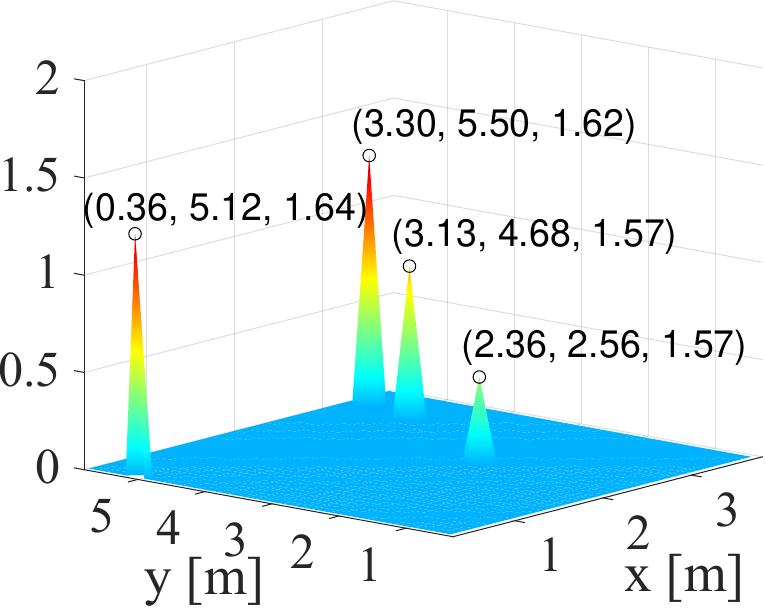}
    \subcaption{}
    \label{a1}
  \end{subfigure}
  \hspace{0.2cm}
  \centering
  \begin{subfigure}[b]{0.17\linewidth}
    \includegraphics[width=1.1\linewidth]{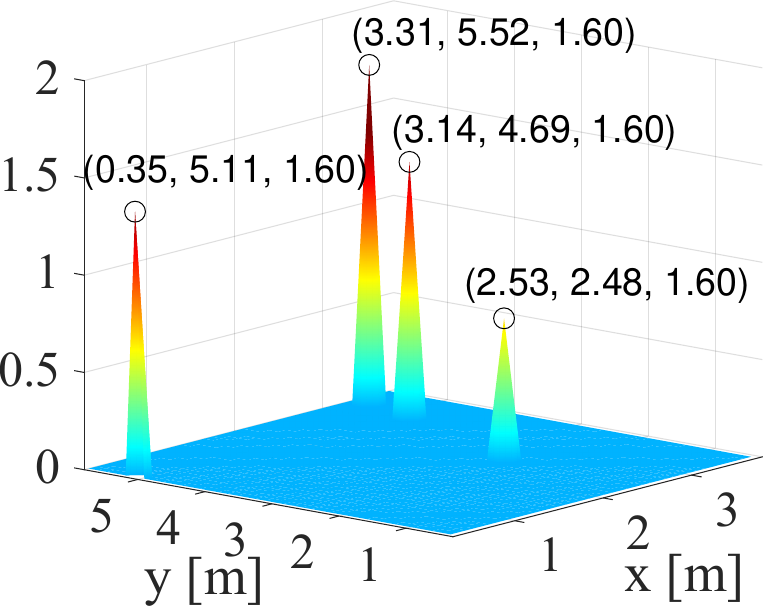}
    \subcaption{}
  \end{subfigure}
  \hspace{0.2cm}
  \begin{subfigure}[b]{0.17\linewidth}
    \includegraphics[width=1.1\linewidth]{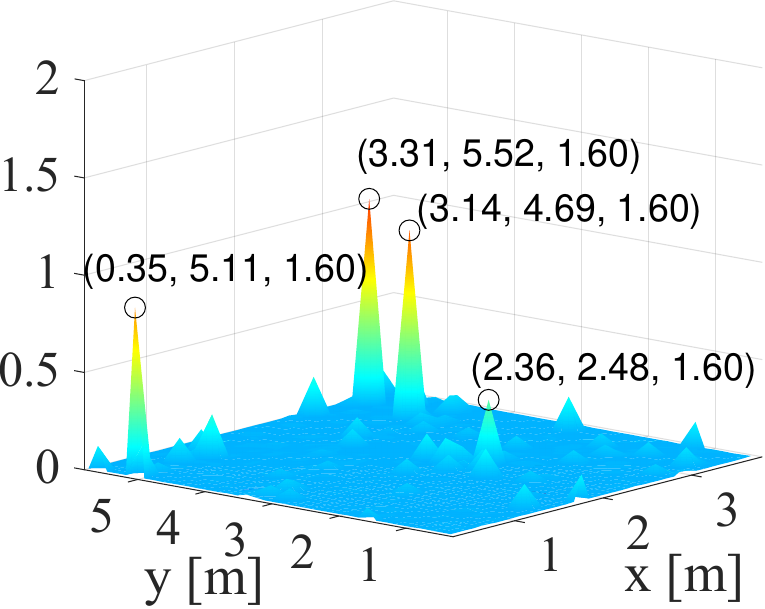}
    \subcaption{}
  \end{subfigure}
  \hspace{0.2cm}
  \begin{subfigure}[b]{0.17\linewidth}
    \includegraphics[width=1.1\linewidth]{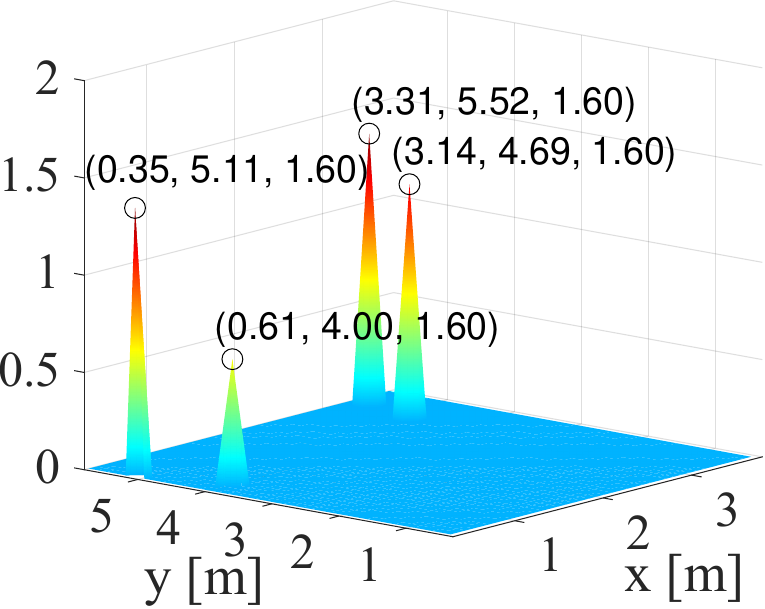}
    \caption{}
  \end{subfigure}
  \hspace{0.2cm}
  \begin{subfigure}[b]{0.17\linewidth}
    \includegraphics[width=1.1\linewidth]{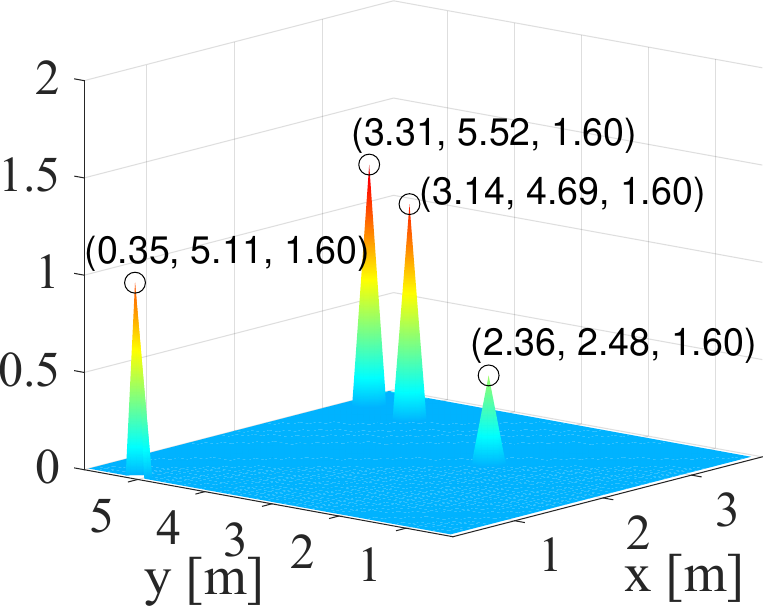}
    \caption{}
  \end{subfigure}
  \par
  \hspace{-0.9cm}
  \begin{subfigure}[b]{0.17\linewidth}
    \includegraphics[width=1.1\linewidth]{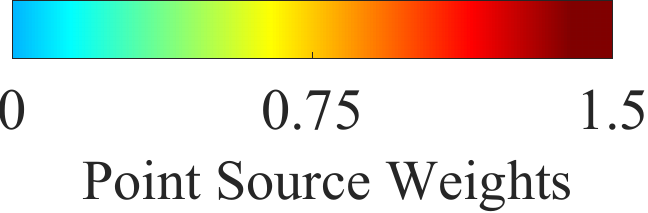}
    \\
    \\
  \end{subfigure}
  \hspace{0.3cm}  
  \begin{subfigure}[b]{0.17\linewidth}
    \includegraphics[width=1.1\linewidth]{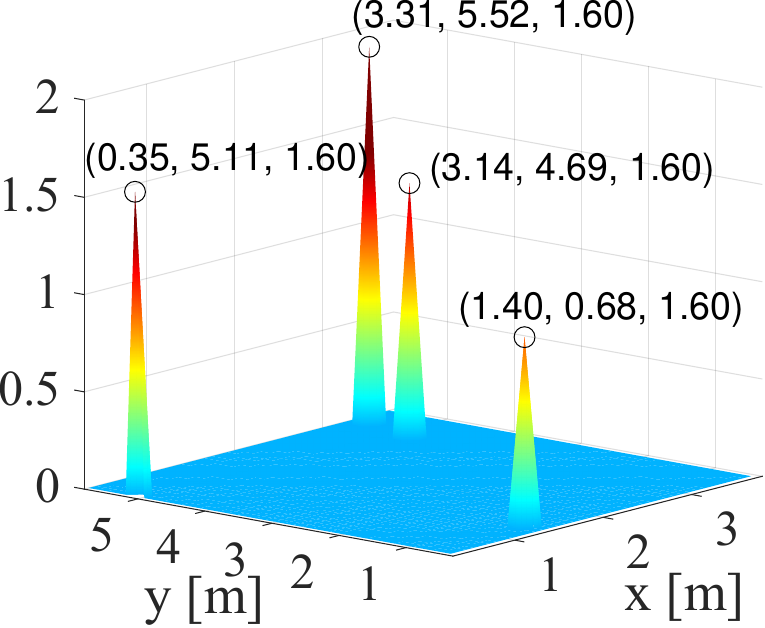}
    \caption{}
  \end{subfigure}
  \hspace{0.2cm} 
  \begin{subfigure}[b]{0.17\linewidth}
    \includegraphics[width=1.1\linewidth]{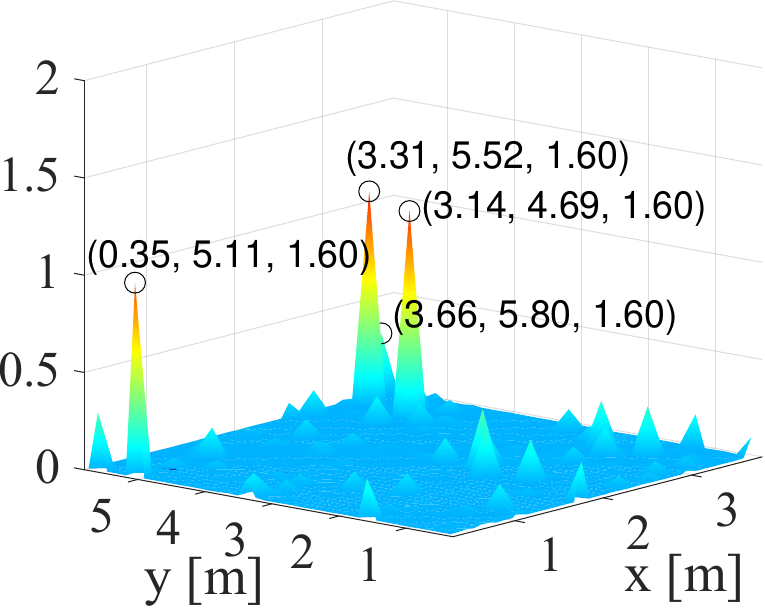}
    \caption{}
  \end{subfigure}
  \hspace{0.2cm} 
  \begin{subfigure}[b]{0.17\linewidth}
    \includegraphics[width=1.1\linewidth]{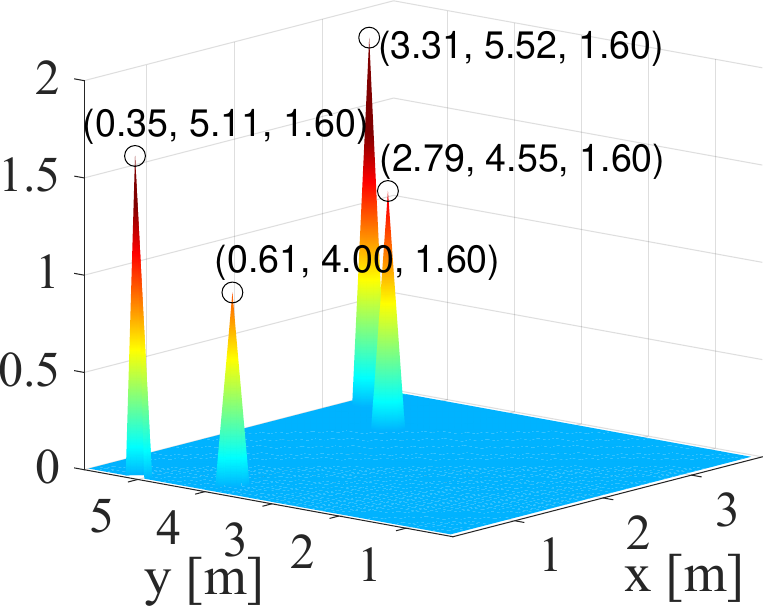}
    \caption{}
  \end{subfigure}
  \hspace{0.2cm} 
  \begin{subfigure}[b]{0.17\linewidth}
    \includegraphics[width=1.1\linewidth]{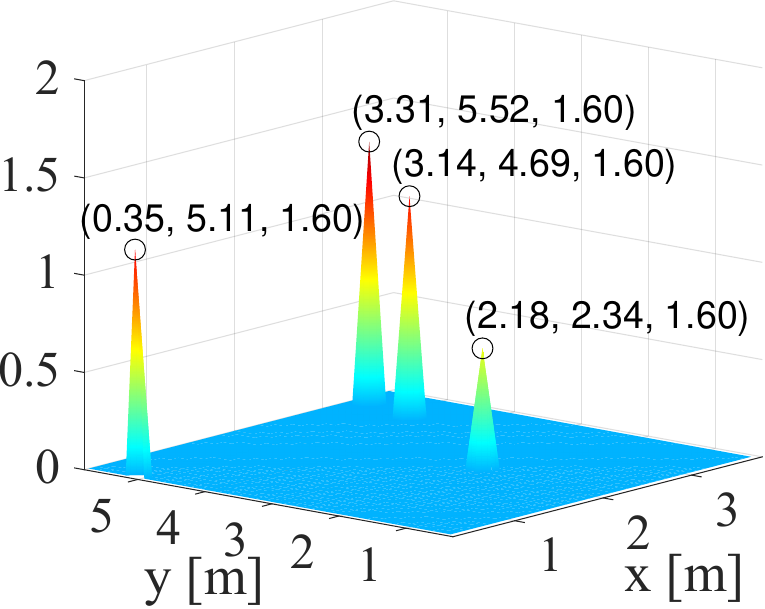}
    \caption{}
  \end{subfigure}
  \caption{Point-source weights at $f = 1000$ Hz in $xy$-plane of (a) ground truth, (b,f) ND-OMP, (c,g) D-IRLS, (d,h) WSPP, and (e,i) proposed, with (a) to (e) for $T_{60}=0.75$ s and (f) to (i) for $T_{60}=1.5$ s.}
\label{figgt}
\vspace{-0.5cm}
\end{figure*}

\begin{figure}[ht]
  \centering
  \centerline{\includegraphics[width=7cm]{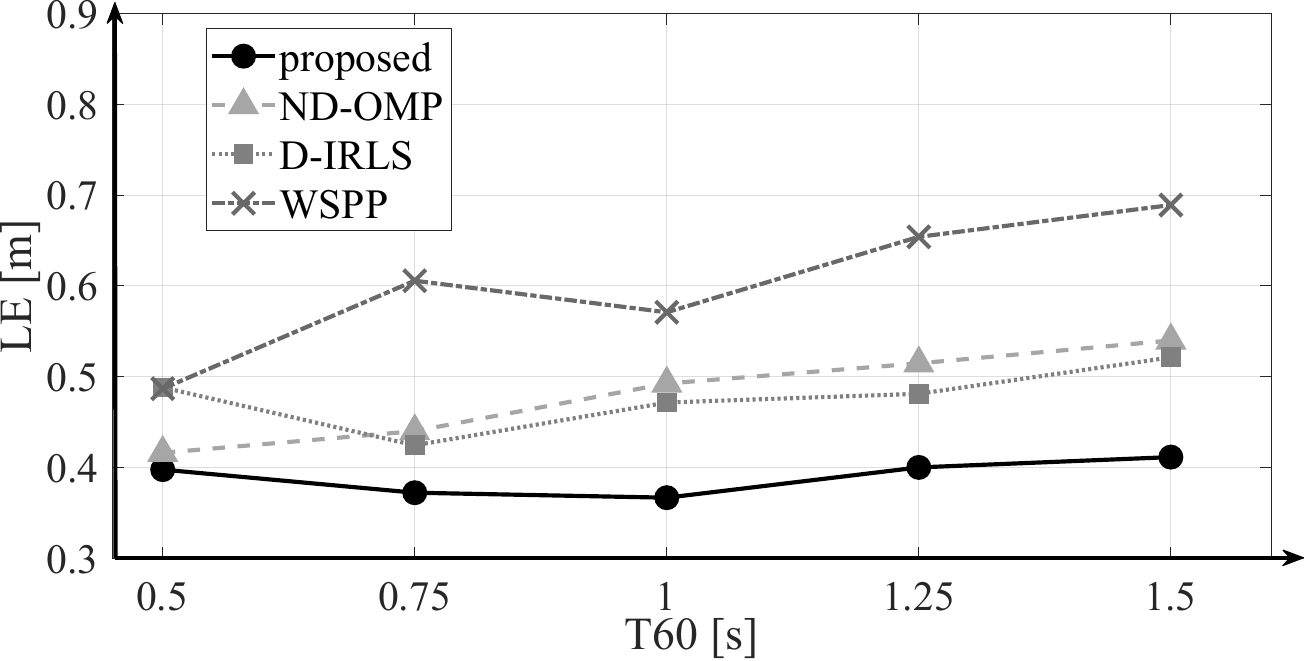}}
 \vspace{-0.1cm}
  \caption{Average localization errors for the four source locations at $f = 1000$ Hz averaged over 100 Monte Carlo tests.}
  \vspace{-0.5cm}
\end{figure}

In this section, we evaluate the proposed method in a simulated reverberant room. To emulate practical scenarios, we focus on a conference room environment with multiple participants conversing. We assume that all sound sources are positioned at approximately 1.6 m height from the ground sitting around the conference table. Therefore, we evaluate the performance of the proposed method for 2D-horizontal plane localization as this minimizes the required microphones for a practical implementation.

We use the RIR generator toolbox \cite{rirgenerator} to simulate a $4.1\times6.2\times3.9$ m reverberant room. We note that while here we model a rectangular shoebox room, our proposed localzation method does not rely on a known room geometry. We position $N=4$ sources within the height range of $1.55\leq z \leq 1.65$ m. Then, we generate the source signals by convolving the image source RIRs with clean speech signals. We select two female and two male speech sources taken from the MS-SNSD dataset \cite{ms-snsd}. The sampling frequency is 16 kHz. The measurements SNR are 30 dB. The STFT parameters are 16384 for the frame length with 50\% overlap. The frequency bin we select is 1 kHz, as $k=18.48$ while the speed of sound $c$ is 340 m/s. Matching the room geometry, we use a uniform rectangular microphone array with $M=106$ microphones, spaced 0.2 m apart, affixed to the perimeter of the walls at a height of $z=1.60$ m. For the reverberant sound field, we select $L=3000$ planewaves and $J=1600$ point-sources (grid points) on the $z=1.60$ m plane, following the 2D localization in a conference room scenario. We assume that the reflection from ceilings and floor are inactive. This implies that we are simulating a 2D sound field using \cite{rirgenerator}.

For comparison, we evaluate three other methods: non-dereverberation by OMP (ND-OMP), WSPP \cite{wspp}, and simultaneous dereverberation and localization by IRLS (D-IRLS). Specifically, ND-OMP estimates the source position directly using OMP without the dereverberation step. WSPP is a method based on planewave decomposition using OMP, eliminating the ambient interference by a linear projection operator. D-IRLS estimates both the direct and reverberant component simultaneously through IRLS, thereby determining $\boldsymbol{\alpha}$ by solving (6). Here, we select $L=70$ planewaves for WSPP and $L=3000$ for D-IRLS.

We first evaluate two specific cases, as shown in Fig. 2. We fix the four source positions as detailed in Fig. 2(a). We select two sets of wall reflection coefficients to compare the performance at different reverberation time ($T_{60}$): [$0.9, 0.93, 0.94, 0.94, 0, 0$] and [$0.99, 0.98,0.98, 0.99, 0, 0$] with the reflection order of image source model as 30, equivalent to 0.75 s and 1.5 s $T_{60}$, respectively.

In Fig. 2, we present the estimated weight vector of point-sources $\boldsymbol{\widetilde{\alpha}}$ compared to the true source weights for both a high and extreme reverberation room. Fig. 2(a) shows the ground truth. Starting with ND-OMP in (b), we see that this method can successfully localize the sources without dereverberation when $T_{60}$ is medium. However, the performance in (f) degrades when $T_{60}$ is high owing to interference from room reflections. Although D-IRLS in (c) and (g) provides a good ability to cancel interference from room reflections, this method is unstable to obtain a sparse solution when estimating $\boldsymbol{\widetilde{\alpha}}$. In figure (d) and (h), the WSPP method is typically effective with a grid-point microphone array \cite{omp_l10n}, but struggles in our setup due to its constrained microphone placements.

The proposed method as observed in (e) and (i) is seen to have the best performance. The two-step approach offers robust localization by combining the advantages of IRLS and OMP. IRLS excels at solving underdetermined systems, so it is not necessary to comply $L \simeq 2 \lceil k r \rceil+1$, where $r$ is the measurement area radius, for optimal dereverberation \cite{vekua2}. This enhances dereverberation performance and prevents overfitting of the reverberant component, allowing flexibility in the number of planewaves to deal with different frequency bins. OMP then provides a strict sparse solution, enhancing source localization and enabling source loudness estimation.

In the second evaluation, we compare the average localization errors (LE) across 100 Monte Carlo test samples for different reverberation scenarios. We randomly placed the four sources inside the conference room at a height range $1.55 {\leq} z {\leq} 1.65$ m. The $T_{60}$ is varied from 0.5 to 1.5 s. We define the LE between the estimated and true source positions to evaluate the performance of the localization methods as:
\begin{equation}
\mathbf{LE} = {\textstyle\frac{1}{N}} {\textstyle \sum_{n=1}^{N}} \|\boldsymbol{\widetilde{y}_n}-\boldsymbol{y}_n\|_2
\end{equation}

The results of Fig. 3 show that the proposed method provides robust source position estimation. As $T_{60}$ increases, the method maintains stable LE values, while the performances of other methods degrade. However, we note that the average LE of the proposed method at $T_{60}=0.5$ s remains relatively high, primarily due to the magnitude differences among different sources. Specifically, if the magnitude of one source is much lower than of the other sources, localizing this particular source becomes challenging because it might be regarded as noise when using sparse recovery methods. The presented results illustrated the robust performance achieved in 2D horizontal-plane localization. However, it is worth to note that the proposed method can be extended to 3D scenarios with 3D point-source and planewave dictionaries, and a 3D microphone array.


\section{CONCLUSION}
\label{sec:print}

In this paper, we have introduced an enhanced method for narrowband source localization in reverberant environments. Based on sparse representation and planewave dereverberation, the two-step approach improves the localization accuracy by estimating the direct and reverberant component separately. The simulation results show that the proposed method can effectively localize multiple sources with a reduced number of measurements, particularly in scenarios with high $T_{60}$. Moreover, our method overcomes the overfitting problem in planewave decomposition, allowing for a more flexible and straightforward determination of planewave quantities. The scope of future work includes resolving the magnitude issue in sparse recovery algorithms and considering wideband scenarios to reduce microphones further.


\vfill\pagebreak
\balance

\bibliographystyle{IEEE}
\bibliography{strings,refs}

\end{document}